\documentclass[aps,twocolumn,preprintnumbers,superscriptaddress,
              showpacs,nofootinbib]{revtex4}
 \newcommand{\PRE}[1]{}       
\usepackage{amsmath,amsfonts,epsfig,graphicx}
\usepackage{multirow}
\usepackage[latin1]{inputenc}
 \usepackage{amssymb}
 \usepackage{verbatim}
\usepackage{float}

\newcommand{\mgme}{{\sc MadGraph/MadEvent}}

\newcommand{\kt}{k_T}
\newcommand{\pt}{p_T}
\newcommand{\pythia}{{\sc Pythia}}

\newcommand{\ie}{{\it i.e.}}

\newcommand\f[2]{\frac{#1}{#2}} 

\begin{document}

\title{Higgs Pair Production: Improved Description by Matrix Element Matching}
 
\author{Qiang Li}
\affiliation{
Department of Physics and State Key Laboratory of Nuclear Physics and Technology, Peking University, Beijing, 100871, China}
\email{qliphy0@pku.edu.cn}

\author{Qi-Shu Yan}
\affiliation{College of Physics Sciences, University of Chinese Academy of Sciences, Beijing 100049, China and
Center for High Energy Physics, Peking University, Beijing 100871, China}
\email{yanqishu@gucas.ac.cn}

\author{Xiaoran Zhao}
\affiliation{College of Physics Sciences, University of Chinese Academy of Sciences, Beijing 100049, China}
\email{zxrlha@gmail.com}

\begin{abstract}
Higgs pair production is crucial for measuring the Higgs boson self-coupling. 
The dominant channel at hadron colliders is gluon fusion via heavy-quark loops. 
We present the results of a fully exclusive simulation of gluon fusion Higgs pair
production based on the matrix elements for $hh + 0, 1$ partons including full
heavy-quark loop dependence, matched to a parton shower. We examine and validate
this new description by comparing it with (a) Higgs Effective Theory predictions, 
(b) exact $hh + 0$-parton sample showered by \pythia, and (c) exact $hh+1$-parton distributions, by 
looking at the most relevant kinematic distributions, such as $p_T^h$, $p_T^{hh}$, $M_{hh}$ spectra, and jet rate as well.
We find that matched samples provide an state-of-the-art accurate exclusive description of the final state. The relevant LHE files for Higgs pair productions at the LHC can be accessed via http://hepfarm02.phy.pku.edu.cn/foswiki/CMS/HH, which can be used for relevant experimental analysis.

\end{abstract}


\maketitle

\section{Introduction}\label{intro}

The discovery of a 125-126 GeV Higgs-like boson~\cite{FGianotti,JIncandela,plb:2012gu,plb:2012gk}, together with the early measurement on its properties~\cite{hp}, forsees a new era in particle physics to understand more in detail the mechanism of electroweak symmetry breaking, in which precision measurement on Higgs couplings will play a very important task, in demand of the forthcoming upgraded LHC and the promising circular or linear $e^+e^-$ collider. In particular, the Higgs self coupling is crucial as it is the only portal to reconstruct and verify the Standard Model (SM) like scalar potential~(see e.g.~\cite{snowmass}):
\begin{equation}
{\cal V}_{\text{SM}}=
-\mu^2\phi^\dagger\phi+\lambda|\phi^\dagger\phi|^2,
\end{equation}
which leads to the triple- and quartic-Higgs couplings
\begin{equation}
g_{HHH}=6\lambda v, \,\, g_{HHHH}=6\lambda,
\end{equation}
In the SM, one has $v=\sqrt{\mu^2/\lambda}$ being the Higgs field vacuum expectation value, and the higgs mass $m_H=\sqrt{2\lambda}v$. Thus the self coupling parameter $\lambda$ is fixed by $m_H$, however, it may not be the case in beyond SM. 

Measuring Higgs pair productions can be sensitive to $g_{HHH}$ and thus $\lambda$ (for recent phenomenology papers, see e.g.~\cite{refhh}). Previous Feasibility studies~\cite{Baglio:2012np,Barger:2013jfa} show that, for example, at the 14TeV LHC with 3000fb$^{-1}$ of data, the SM Higgs pair process can be observed with high significance and the trilinear coupling can be measured within 40\% accuracy.
At the LHC, Higgs boson pair production mainly proceeds via gluon
fusion (GF) induced by heavy-quark triangle and box loops, with the former being sensitive
to the higgs trilinear couplings. In the large $M_t$ limit, one can integrate out the top quark filed, resulting, to a good approximation, in a simple, non-renormalizable effective field theory (HEFT) ~\cite{Dawson:1998py}, 
\begin{equation}\label{diheft}
{\cal L}_{\text{eff}}=
\f{\alpha_s}{12\pi}G^a_{\mu\nu}G^{\mu\nu,a}\log{\left(1+\f{H}{v}\right)},
\end{equation}
with $G^{\mu \nu,a}$ being the QCD field tensor. 

The leading order (LO) exact cross section has been evaluated long ago~\cite{Glover:1987nx,Eboli:1987dy,Plehn:1996wb}.
The next-to-leading order (NLO) QCD corrections have also been calculated
in Ref.~\cite{Dawson:1998py}, within the HEFT, resulting a K factor close to 2.
The next-to-next-to-leading order (NNLO) QCD results are presented in Ref.~\cite{deFlorian:2013uza,deFlorian:2013jea},
again within the HEFT, finding a further increase of a factor of $\sim 1.2$ over the NLO one. The top quark mass effect at NLO has also been studied in Ref.~\cite{Grigo:2013rya}, which shows a good accuracy at ${\cal O} ( 10\%)$ can be achieved if the LO result is normalized taking into account the top quark mass.

On the other hand, in experimental analyses, it is crucial to get as precise predictions as possible for exclusive observables, such as the transverse momentum of the higgs pair, $p_T^{hh}$, and jet rate as well. As well known, the differential distributions predicted by HEFT is not trustable due to lost loop information, which can be easily seen if compared e.g. the HEFT $M_{hh}$ and $p_T^{h}$ spectrum with the exact loop results as shown in Figs.~\ref{mhh}-\ref{pth}. One can also build in the exact loop induced matrix elements (ME) from e.g. MadLoop~\cite{Hirschi:2011pa} within MadEvent~\cite{Maltoni:2002qb} for showering and event generating, as done in~\cite{Barger:2013jfa}. However, beyond that, there is no available fully exclusive prediction yet so far. The reason is that one needs to compromise between the validity of HEFT and the complexity of higher loop calculations. And this is the current Monte-Carlo tool status for LHC Higgs pair productions.

It is however possible to get full exclusive control
at hadron level on the complex event topology at the LHC, while still
reaching approximately Next-to-leading Logarithms (NLL) accuracy, with the help of recent
sophisticated matching methods between ME and parton
showers (PS)~\cite{Catani:2001cc,Alwall:2007fs}. In PS programs, QCD radiation is generated in the collinear and soft approximation, using Markov chain techniques based on Sudakov form factors. Hard and widely separated
jets are thus poorly described in this approach. On the other hand,
tree-level fixed order amplitudes can provide reliable predictions in
the hard region, while failing in the collinear and soft limits. To
combine both descriptions and avoid double counting or gaps between
samples with different jet multiplicity, an appropriate matching method is
required. Several algorithms have been proposed over the years: the
CKKW method, based on a shower veto and therefore on event
re-weighting~\cite{Catani:2001cc} and MLM
schemes, based on event rejection~\cite{Mangano:2006rw,
  Alwall:2007fs}. 

In this work, we report on  the first matched simulation of Higgs pair
production via GF in the SM that retains the full kinematic
dependence on the heavy-quark loops. 

The paper is organized as follows.
We begin by describing our methodology. Then we present our results for
the SM Higgs pair productions. We show that the matching procedure provides reliable
results at the LHC and that the effects from massive quark loops
can be significant. Finally we conclude in the last section.

\section{Method}

Our study is based on the $\kt$-MLM and shower-$\kt$ matching
schemes~\cite{Alwall:2007fs,Alwall:2008qv}, implemented in \mgme~\cite{Alwall:2007st}, 
interfaced with \pythia~6.4~\cite{Sjostrand:2006za} for parton shower and hadronization. As explained more in detail in our previous works for single Higgs case~\cite{Alwall:2011cy} , 
we find it convenient to include the effects of the heavy-quark loop by simply 
reweighting the events generated via tree-level HEFT amplitudes.

In short, we implement di-Higgs Gluon effective interactions Eq.~(\ref{diheft}) into \mgme, and then generate parton level events for  $hh+0 ,\,1$ partons in this model. Before passing them to
the PS program, events are reweigthed by the ratio of full
one-loop amplitudes over the HEFT ones, $r= |{\cal M}_{\rm LOOP}|^2/|{\cal M}_{\rm HEFT}|^2$, where $|{\cal M}_{\rm LOOP}|^2$ represents the full one-loop amplitude got by FeynArts~3.5 \cite{Hahn:2000kx},
FormCalc~5.3 \cite{Hahn:1998yk} and LoopTools-2.5 package \cite{Hahn:1998yk}. The reweighted parton-level events are unweighted, passed through \pythia\ and matched using the shower-$k_T$ scheme. All steps are automatic. To validate the matching procedure,  the effect of changing the matching cutoff parameters
such as $Q^{\rm jet}_{\rm min}$ and $Q^{\rm  ME}_{\rm min}$ in the shower-$\kt$ matching
schemes~\cite{Alwall:2007fs,Alwall:2008qv} on several distributions, including the $n \to n-1$ differential jet rates have been extensively assessed.

Finally, we recall that even though matrix elements for up to one final states partons are included in the simulation, the accuracy of the overall normalization of the inclusive sample is only leading order, exactly as in a purely parton-shower result. 

In the following, for convenience, we call the above achieved results as Loop with matching, which will be compared with HEFT with matching, exact $hh+1$ parton level predictions from FeynArts/FormCalc/LoopTools, and exact $hh + 0$ parton sample showered predictions which we get by interfacing MadLoop with our standalone MC generator for event generating and showering.

\section{ SM Higgs Pair production}

\begin{figure}
\includegraphics[width=8.5cm]{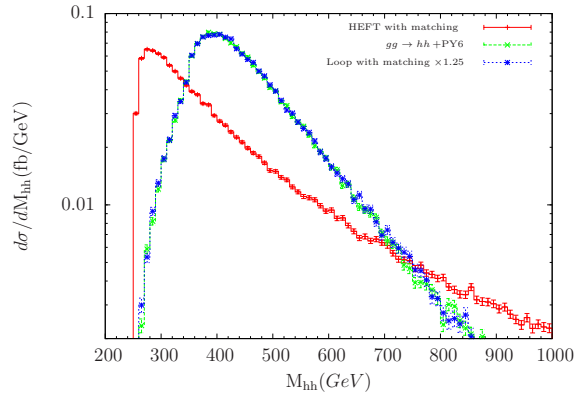}
\caption{$M_{hh}$ distributions in Higgs pair gluon fusion production at the 14TeV LHC.}
\label{mhh}
\end{figure}

\begin{figure}
\includegraphics[width=8.5cm]{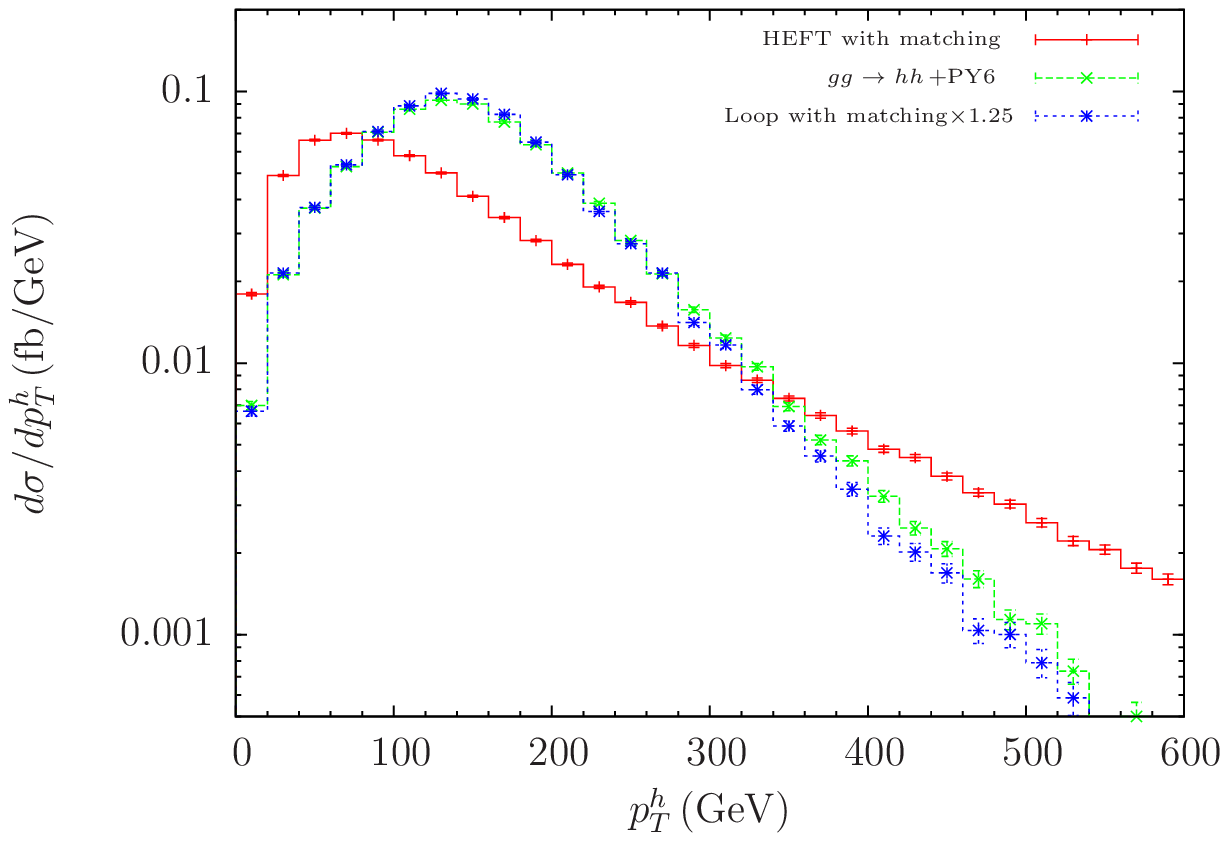}
\includegraphics[width=8.5cm]{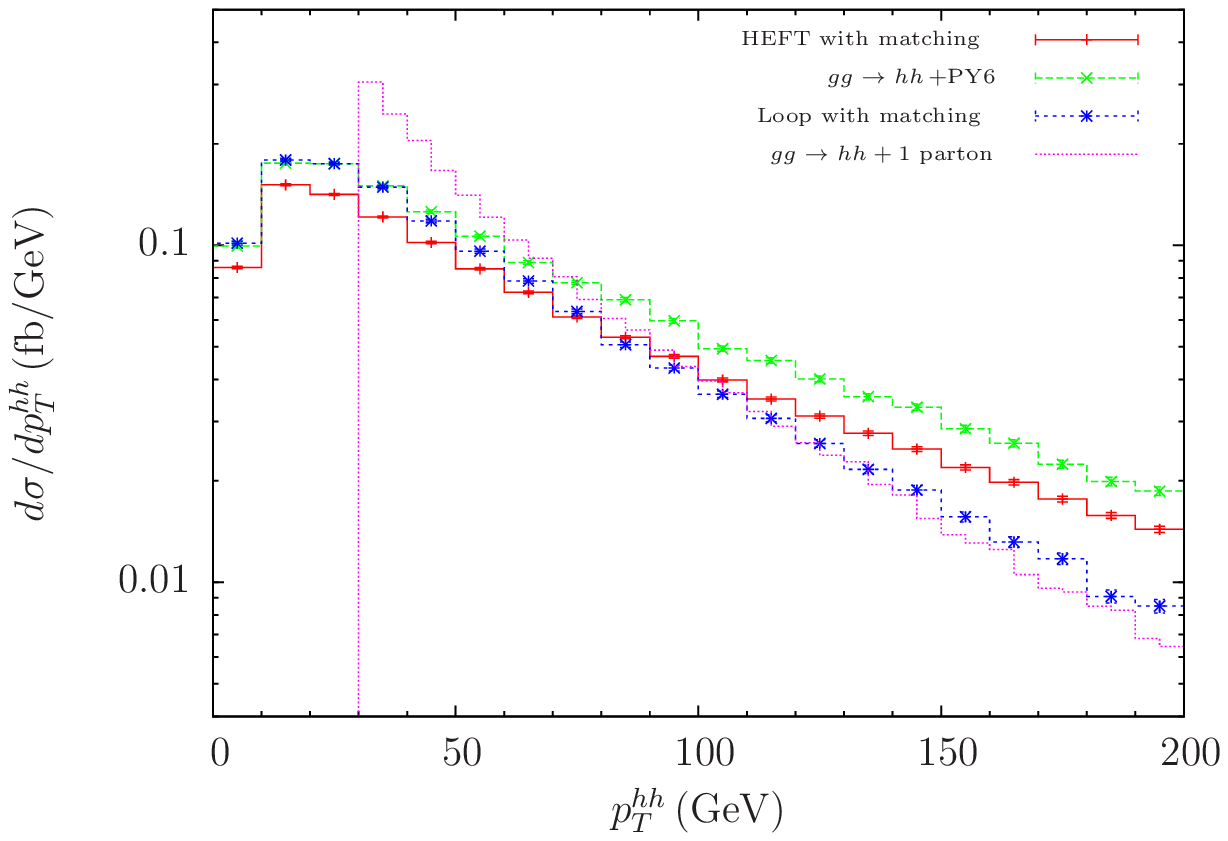}
\caption{Single and di-Higgs  $\pt$ distributions for
Higgs pair gluon fusion production at the 14 TeV LHC.}
\label{pth}
\end{figure}

\begin{figure}
\includegraphics[width=8.5cm]{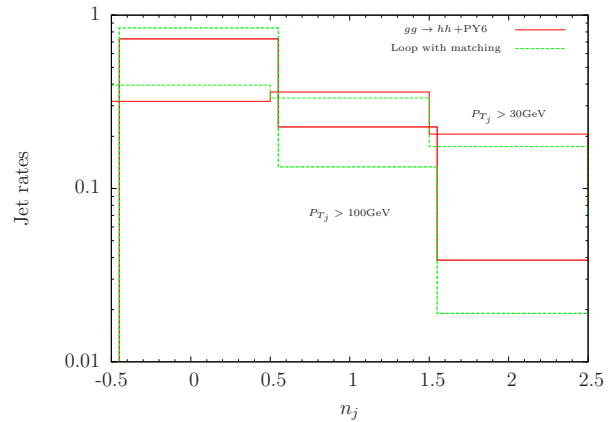}
\caption{Jet rates for Higgs pair gluon fusion production at the 14TeV LHC.}
\label{hhjr}
\end{figure}

\begin{figure}
\includegraphics[width=8.5cm]{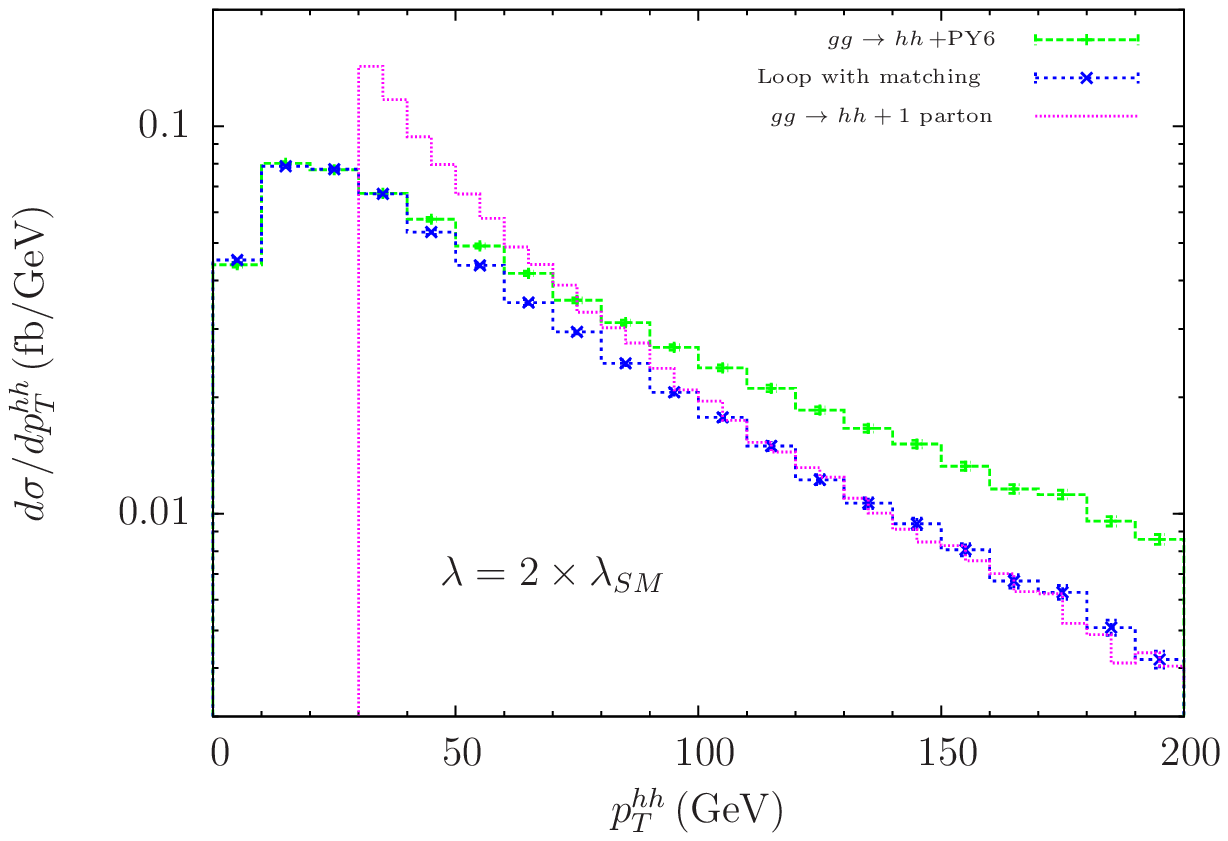}
\includegraphics[width=8.5cm]{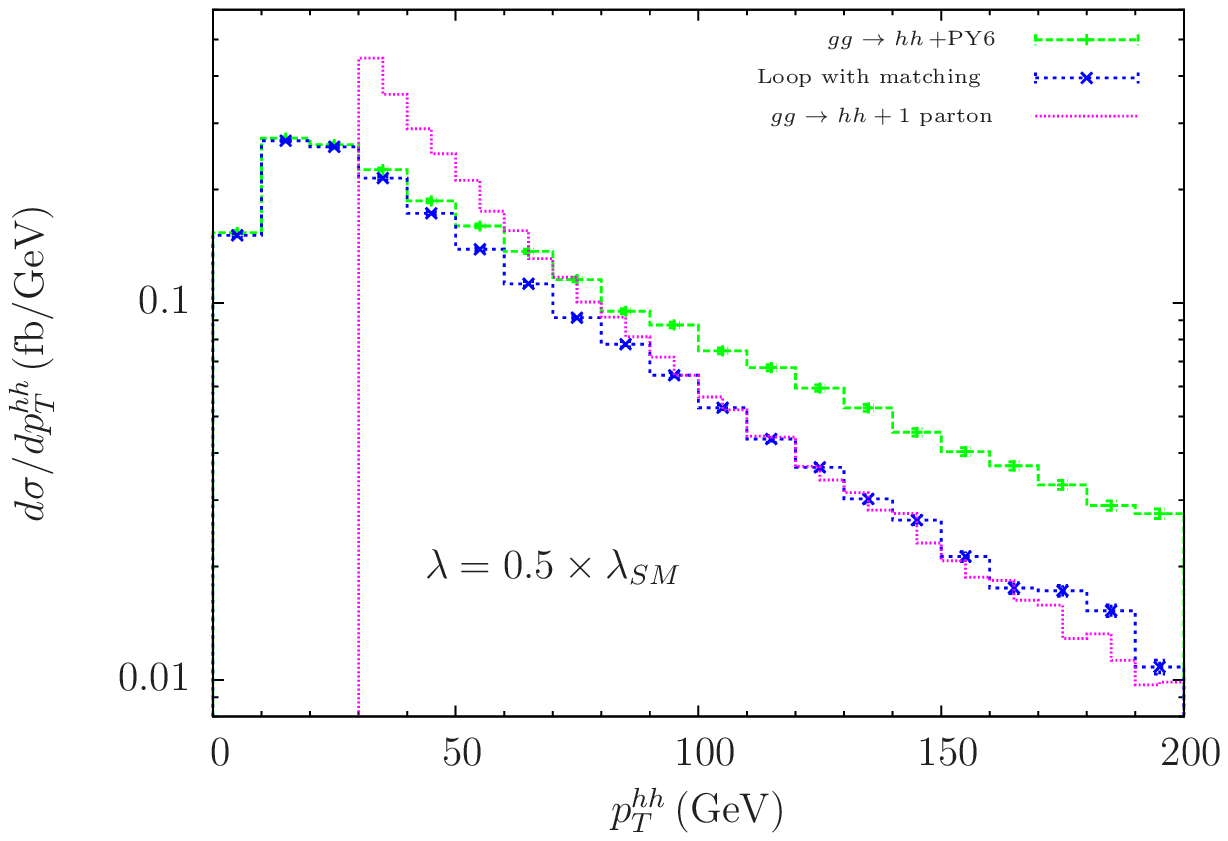}
\caption{$p_T^{hh}$ distributions for
Higgs pair gluon fusion production at the 14 TeV LHC, with the trilinear Higgs parameter $\lambda$ scaled from the SM value by a factor of 2 and 0.5.} 
\label{pthlambda}
\end{figure}

To illustrate the results of our simulations for Higgs pair production via GF at the 14TeV LHC, 
we show a few relevant observables in Figs.~\ref{mhh}-\ref{pthlambda}. We define jets via the $\kt$ algorithm, with the distance measure between
parton $i$ and beam $B$, or partons $i$ and $j$ as $k_T^{i,B}\equiv
p_T^i$, $k_T^{i,j}\equiv \min{\left(p_T^i, p_T^j\right)}
\sqrt{2(\cosh\Delta y_{ij}-\cos\Delta\phi_{ij})}/D$.  
Here $y$ is the rapidity and $\phi$ is the azimuthal angle around the beam
direction. The resolution parameter is set to $D = 0.7$. Jets are required to satisfy $|\eta_{j}|< 4.5$ and $p_T^{j}>30\,{\rm GeV}$.
For sake of simplicity, we adopt Yukawa couplings corresponding to the
pole masses, \ie,  for the top quark  $m_t=173$\,GeV and for the bottom-quark mass $m_b=4.6$\,GeV. Other quark masses are neglected. 
Throughout our calculation, we set $m_H=126$\,GeV, and adopt the CTEQ6L1 parton distribution functions
(PDFs)~\cite{Pumplin:2002vw} with the core process renormalization and
the factorization scales $\mu_r=\mu_f$ to parton center of mass energy, $\sqrt{\hat{s}}$.
For the matching performed in \mgme, as mentioned above, the 
shower $\kt$-MLM scheme is chosen, with $Q^{\rm ME}_{\rm min}=Q^{\rm jet}_{\rm min}=40$\,GeV.

In Figs.~\ref{mhh} we show Higgs pair invariant mass distribution for Standard Model Higgs pair GF production at the 14TeV LHC.  
We compare matched results in the HEFT theory and in the full theory
(Loop) with \pythia.  We also include the predictions from exact $hh + 0$ parton showered with \pythia. 
The curves are all normalized to their own predictions, except the Loop with matching one which is scaled by a factor of 1.25.~\footnote{The exact loop cross section reads 16.1fb and the Loop one 12.9fb} The HEFT doesn't not describe the top quark effects well as expected, while the other two agree well each other as should be.
 
In Figs.~\ref{pth}, we show single and di-Higgs $\pt$
distributions for Standard Model Higgs GF production at
the 14TeV LHC.~\footnote{Note in the $p_T^{hh}$ plot, we don't scale the Loop with matching curve but keep its own normalization to show explicitly the agreement with exact $hh + 1$ parton level curve.} In the $p_T^h$ plot, one can again see the HEFT does not describe well the behavior. Instead, as expected, loop effects show
a softening of the Higgs $\pt$, especially at quite high $\pt$. The Loop
with matching curve agrees well with exact $hh + 0$ parton showered result, except being 
a bit soft at high $\pt$ tail, which is due to the softening effect from exact $hh + 1$ parton
contributions included. 

The $p_T^{hh}$ distributions essentially reflects the accompanying jet information
and thus are quite sensitive to the loop effects. As one can see, the HEFT and exact $hh + 0$ parton showered samples
predict a too hard tail, although at low $p_T^{hh}$ they agree well with Loop with matching curve.
This is again due to the softening effect from exact $hh + 1$ parton contributions, which can be seen from the 
agreement between Loop with macthing and the exact $hh + 1$ parton level curves. Note at very high $p_T^{hh}$,
the Loop with matching prediction is a bit harder than the $hh + 1$ one, because of additional jet radiation from shower. 

Fig.~\ref{hhjr} shows the corresponding jet rates for different minimum jet $p_T^j$
of 30 and 100 GeV. As is readily seen from the figure, the
effect of properly including loop effects is significant already at the order of 10-20\% with
a jet $p_T^j$ cutoff at 30 GeV, and increasingly important for larger
cutoff values. This immediately translates to the effect of e.g. a jet veto
with a given $p_T^j$ cutoff for the veto.

Finally, in In Fig.~\ref{pthlambda} we show the $p_T^{hh}$ distributions for the cases
of trilinear Higgs parameter $\lambda$ scaled from the SM value by a factor of 2 and 0.5, with similar behaviour as the above SM case,
which proves our framework also works here.

\section{Conclusions}

In summary, we have presented the first fully exclusive simulation
of gluon fusion inclusive Higgs pair production based on the exact one-loop
matrix elements for $hh+0,\,1$ partons, matched to \pythia\
parton showers using shower-$k_T$ matching schemes implemented in \mgme.
We have compared the loop reweighted matched results 
with the corresponding  HEFT results, exact $hh + 0$ parton showered results, and, when possible,  with exact $hh + 1$ parton level predictions. We have studied the most relevant kinematic distributions, such as Higgs pair $\pt$ spectra and jet rates.
Our results highlight the importance of a complete loop calculation at large $\pt$ for a standard model Higgs.
Such improved simulations might be particularly relevant in searches performed via multivariate analysis  techniques where details about the kinematic distributions of the Higgs decay products and accompanying jets can have significant impact on the results.

\acknowledgments
This work is supported in part by the National Natural Science Foundation of China, under Grants No. 11175251 and 11205008.

\bibliographystyle{ieeetr}
\bibliography{h}
\end{document}